\documentclass[aps,pra,twocolumn,groupedaddress]{revtex4}
\usepackage{graphicx}

\begin{document}


\title{Dynamical decoupling induced renormalization of the non-Markovian dynamics}


\author{Pochung Chen}
\affiliation{Department of Physics, National Tsing-Hua University,
  Hsinchu 100, Taiwan}


\date{\today}

\begin{abstract}
In this work we develop a numerical framework to investigate the renormalization
of the non-Markovian dynamics of an open quantum system to which
dynamical decoupling is applied. 
We utilize a non-Markovian master equation which is derived from 
the non-Markovian quantum trajectories formalism.
It contains incoherent Markovian dynamics and coherent Schr\"odinger dynamics 
as its limiting cases and is capable of capture the transition between them.
We have performed comprehensive simulations for the cases in which the system 
is either driven by the Ornstein-Uhlenbeck noise or
or is described by the spin-boson model.
The renormalized dynamics under bang-bang control and 
continuous dynamical decoupling are simulated.
Our results indicate that the renormalization of the non-Markovian
dynamics depends crucially on the spectral density of the environment
and the envelop of the decoupling pulses.
The framework developed in this work hence provides an unified approach
to investigate the efficiency of realistic decoupling pulses.
This work also opens a way to further optimize the decoupling via pulse shaping.

\end{abstract}


\maketitle

\section{Introduction}

It is known that dynamical decoupling\cite{viola:2733,PZ99,viola:4888,viola:2417}
is one of the essential control tools in fighting the decoherence of an open
quantum system. In recent years 
many methods have been developed to design the decoupling pulses.
Those methods are based on the group structure or the geometric
perspective of the system environment interaction
\cite{viola:4888,viola:2417,viola:037901, chen:022343, byrd:2002,viola:060502}.
The effect of a general environment, however, is usually neglected.
Dynamical decoupling can be divided into several categories.
It can be deterministic and periodic\cite{viola:2733,sengupta:037202},  
deterministic and concatenated\cite{khodjasteh:180501}, or
random\cite{facchi:032314,viola:060502}.
The strength of the pulses can be unbounded, which is usually termed bang-bang decoupling,  or bounded \cite{viola:037901,chen:022343}.
Bounded control uses more realistic control resources and is more tolerant
against implementation errors.
Both unbounded and bounded control attain the desired decoupling in the 
ideal limit where the period of a single decoupling cycle approaches zero.
In practice, however, 
the duration of the cycle is finite and there will always be residual decoherence.
To the best of our knowledge it is difficult to calculate the exact residual decoherence,
especially when the effect of a general environment are taken into account.
For a general configuration the amount of residual decoherence 
might depend on the shape of the 
decoupling pulses as well as the spectral density of the environment.
For example, it has been argued that the bang-bang decoupling might be
more efficient when the environment is characterized by the $1/f$ noise
in stead of an Ohmic bosonic environment\cite{shiokawa:030302, faoro:117905}.
Recently the effect of a general environment is studied, 
but only unbounded control are investigated\cite{shiokawa_2005}.
In other words, a comprehensive study of the residual decoherence
of a system, which is coupled to a general environment and to which
a general dynamical decoupling is applied, is still missing.
This is mostly due to the lack of a convenient tool.
An unified framework to investigate these effects is thus much needed.

In this work we develop an unified numerical framework
to simulate the renormalization of non-Markovian dynamics 
of a system to which decoupling pulses are applied.
It is capable of handling both unbounded and bounded pulses.
It is also able to capture the effect of a general environment.
Such a framework can help us to study the efficiency of a prescribed 
decoupling pulse in realistic situations. It might also open new avenues
to optimize the decoupling pulses.
To see what are the necessary ingredients of such a framework,
recall that a quantum system without decoherence
obeys Schr\"odinger equation, resulting a coherent, hence non-Markovian, dynamics.
On the other hand, long time dynamics of an open quantum system
in high temperature regime is typically Markovian\cite{caldeira:83}. 
A successful decoherence control should renormalize the dynamics,
moving the system dynamics away from Markovian regime and 
move it into the non-Markovian regime. 
The residual decoherence, however, prevent us from simply using the Schr\"odinger equation to describe the renormalized dynamics.
It is thus essential that such a framework can
describe both the Markovian and non-Markovian dynamics for an open quantum 
system coupled to a general environment. 
Also, for numerical purpose
a nonconvoluted master equation is highly desirable 
comparing to an integral-differential equation.

The nonconvoluted non-Markovian master equation \cite{yu:91,vega:023812} 
derived from the non-Markovian quantum trajectories \cite{diosi:1699} 
formalism naturally provides such a framework.
Markovian dynamics and Schr\"odinger dynamics can be seen as two 
limiting cases of the formalism. 
The formalism has been used to study the decoherence of the open
quantum system in the non-Markovian regime \cite{PhysRevB.66.193306,yu:165322}
as well as the non-Markovian dynamics of a two-level atom immersed in a photonic 
band-gap material\cite{vega:023812}. In some cases the stochastic non-Markovian
quantum trajectories can be formally averaged, resulting an exact non-Markovian 
master equation. In general, when the decoupling are applied to the system,
an exact non-Markovian master equation cannot be formed\cite{yu:91}. 
But an approximated master equation can always be derived in the nearly Markovian 
limit\cite{yu:91} or weak coupling limit\cite{vega:023812}.
It has been shown that the resulting non-Markovian master equation
is valid in a wide range of parameter regime\cite{yu:91,vega:023812}.
It is this nonconvoluted master equation that will be taken as the starting
point of our investigation of the renormalization of the non-Markovian dynamics.

The organization of the manuscript is as follows.
In Section \ref{sec:nm} we briefly review the non-Markovian
quantum trajectory formalism and the resulting non-Markovian master equation.
In section \ref{sec:dd} we derive the non-Markovian master equations when
the decoupling pulses are applied.
In Section \ref{sec:ou} we simulate the renormalized dynamics after decoupling
of a system which is driven by the Ornstein-Uhlenbech noise while 
in Section \ref{sec:sb} we study the renormalized dynamics of the spin-boson model. 
Three representative spectral densities are used. 
Bang-bang control pulses and continuous decoupling pulses are used in each case.
The conclusion in Section \ref{sec:conclusion} briefly summarizes our findings. 

\section{Non-Markovian quantum trajectories and master equations}
\label{sec:nm}
The total Hamiltonian of our model of the open quantum system takes the form
$  H_{tot}=H_{sys}+H_{env}+H_{int}$ ,
where 
$  H_{sys}=H_0+H_c(t)$
represents the system Hamiltonian which has been split into the original
Hamiltonian $H_0$ and the dynamical decoupling control Hamiltonian $H_c$.
$  H_{env}=\sum_\lambda \omega_\lambda a^\dagger_\lambda a_\lambda $
represents the bosonic Hamiltonian of the environment and
  $H_{int}
  =\left( \sum_\lambda L g^*_\lambda a^\dagger_\lambda
  +L^\dagger  g_\lambda a_\lambda \right)$
represents the interaction between system and environment.
In our simulations we have set $H_0=0$ since we mainly focus on quantum memory
In this work.

It has been shown that one can derive linear stochastic equations
governing the dynamics of the system under the influence of the environment
\cite{diosi:1699}.
The stochastic Schr\"odinger equation can be obtained by expressing the
total state $|\Psi(t)\rangle$ in terms of the Bargmann coherent state of the
environment 
\begin{equation}
  |\Psi(t)\rangle=\int \frac{d^2 z}{\pi} e^{-|z|^2} |\psi(z^*,t)\rangle|z\rangle,
\end{equation}
where $|z_\lambda\rangle \equiv exp(z_\lambda a^\dagger_\lambda)|0\rangle$ 
and $|z\rangle=|z_1\rangle|z_2\rangle\cdots|z_\lambda\rangle\cdots$.
The resulting equation of $|\psi(z^*,t)\rangle$ reads
\begin{eqnarray}
  \label{eq:NM_QSD_1}
  \frac{d}{dt} |\psi(z^*,t)\rangle
  &=& \nonumber \left[ -iH_{sys}
  + L \eta^*(t)  \right] |\psi(z^*,t)\rangle  \\
  &-& L^\dagger \int_0^t ds \alpha(t,s) \frac{\delta |\psi(z^*,t)\rangle}{\delta \eta^*(s)},
\end{eqnarray}
where 
  $\eta^*(t)=-i \sum_\lambda g^*_\lambda z^*_\lambda e^{i\omega_\lambda t}$
and 
\begin{equation}
  \label{eq:alpha}
  \alpha(t,s)
  =\sum_\lambda |g_\lambda|^2 e^{-i\omega_\lambda(t-s)}
  =\int_0^\infty d\omega J(\omega) e^{-i\omega (t-s)}.
\end{equation}
Using this representation the reduced density matrix of the system
can be expressed as
\begin{equation}
  \rho(t)=\int \frac{d^2z}{\pi} e^{-|z|^2} |\psi(z^*,t)\rangle\langle\psi(z^*,t)|.
\end{equation}
If one choose to use Monte-Carlo integration to evaluate this expression,
taking over the Gaussian distribution 
\begin{equation}
  \mathcal{M} [\dots]
  =\int\frac{d^2 z}{\pi} [\dots]
\end{equation}
of $z-$vectors, then $\eta^*(t)$ can be interpreted as stochastic variables
with following statistical properties
\begin{equation}
  \mathcal{M}[\eta(t)]=\mathcal{M}[\eta(t)\eta(s)]=0;
\end{equation}
\begin{equation}
  \mathcal{M}[\eta(t)\eta^*(s)]=\alpha(t,s).
\end{equation}
Under this interpretation, Eq.(\ref{eq:NM_QSD_1}) becomes a stochastic
Schr\"odinger equation on the system. However it is still difficult to
treat Eq.(\ref{eq:NM_QSD_1}) due the the functional derivative with
respect to the noise $\eta^*(s)$ under the memory integral.
A nonconvoluted stochastic equation can be obtained by proposing
the Ansatz
\begin{equation}
  \frac{\delta |\psi(z^*,t)\rangle}{\delta \eta^*(s)}=O(t,s,\eta(t))|\psi(z^*,t)\rangle,
\end{equation}
where $O(t,s,\eta(t))$ is a linear operator that has to be determined for each
case. With this replacement the stochastic equation becomes
\begin{equation}
  \label{eq:NM_QSD_2}
  \frac{d}{dt} |\psi(z^*,t)\rangle
  =\nonumber \left [-iH_{sys}
  + L \eta^*(t) -L^\dagger \bar{O}(t,\eta^*) \right] |\psi(z^*,t)\rangle,
\end{equation}
where
  $\bar{O}(t,\eta^*)=\int_0^t ds \alpha(t,s)O(t,s,\eta^*)$.
For some cases it is possible to construct exactly the $O$ operator.
When it is difficult to identify the $O$ operator, it is still possible to derive
an approximate $O$ operator by perturbation in terms of the coupling 
parameter \cite{vega:023812} or 
the environmental correlation time\cite{yu:91}, If the exact 
or approximate $O$ operator is independent of the noise $\eta^*$,
a master equation can be easily derived and takes the form:
\begin{equation}
  \frac{d\rho(t)}{dt}
  =-i[H_{sys},\rho]
  +[L,\rho(t)\bar{O}^\dagger(t)]+[\bar{O}(t)\rho(t),L^\dagger].
\end{equation}
Since we are interested in the renormalization of the non-Markovian
dynamics we take the weak coupling limit in stead of the nearly Markovian limit.
In this limit
\begin{equation}
  O(t,s,\eta^*)\approx U(t-s) L U^\dagger(t-s),
\end{equation}
where
\begin{equation}
  U(t)=\mathcal{T} e^{-i \int_0^t du H_{sys}(u)}.
\end{equation}
Hence
\begin{equation}
  \bar{O}(t)= \int_0^t ds \alpha(t,s) U(t-s) L U^\dagger(t-s).
\end{equation}
In the interaction picture with respect to the system Hamiltonian
  $\tilde{\rho}(t)=U^\dagger(t) \rho(t) U(t)$,
the master equation becomes
\begin{eqnarray}
  \label{eq:NM_ME_1}
  \frac{d\tilde{\rho}(t)}{dt} 
  &=& \nonumber 
  \left[\tilde{L}(t),\tilde{\rho}(t) \int_0^t ds \alpha^*(t,s) \tilde{L}^\dagger(s)\right] \\
  &+& \left[\int_0^t ds \alpha(t,s) \tilde{L}(s) \tilde{\rho}(t), \tilde{L}^\dagger(t)\right],
\end{eqnarray}
where
  $\tilde{L}(t)=U^\dagger(t) L U(t).$
Note that when $\alpha(t,s)\rightarrow \delta(t-s)$ the master equation
reduces to the standard Markovian Lindblad master equation in Heisenberg
picture, i.e. 
\begin{equation}
  \frac{d\tilde{\rho}(t)}{dt}
  =[\tilde{L}(t),\tilde{\rho}(t)\tilde{L}^\dagger(t)]
  +[\tilde{L}(t)\tilde{\rho}(t),\tilde{L}^\dagger(t)].
\end{equation}

\section{Dynamical decoupling}
\label{sec:dd}
In the language of standard periodic dynamical decoupling one usually 
starts from the total Hamiltonian
$H_{tot}=\sum_\gamma S_\gamma \otimes B_\gamma$. A periodic control
Hamiltonian $H_c(t)$ with period $T_c$ is introduced to
decouple the system from the environment. Using Magnus expansion\cite{magnus54}
the average Hamiltonian governing the stroboscopic dynamics has the form
\begin{equation}
  \bar{H}_{tot}=\frac{1}{T_c}\int_0^{T_c} ds 
  \left[ U^\dagger_c(s) \sum_\gamma S_\gamma \otimes U_c(t) \right] B_\gamma.
\end{equation}
A properly designed control Hamiltonian can generate an $U_c(t)$ which
results in a zero $\bar{H}_{tot}$. 
Many methods have been developed to design the decoupling pulses
\cite{viola:4888,viola:2417,viola:037901, chen:022343,byrd:2002,viola:060502}.
In this work we focus on investigating the renormalized dynamics
when a decoupling pulses based on those methods are applied to the system. 
It should be noted, however, that it is also possible to design decoupling
pulses based on Eq.(\ref{eq:NM_ME_1}).
We sketch the procedure as follows.
First decomposes $L$ into $L=L_A+i L_B$ where both $L_A$ and $L_B$ are Hermitian. 
It is then easy to verify that the decoupling condition (for the case of quantum memory) is
\begin{equation}
  \int_0^{T_c} ds \alpha(t,s) \tilde{L}_A(s) =\lambda_A, \; 
  \int_0^{T_c} ds \alpha(t,s) \tilde{L}_B(s) =\lambda_B,
\end{equation}
where $\lambda_A$ and $\lambda_B$ are real numbers.
When this condition is satisfied, the master equation turns into
\begin{eqnarray}
  \left. \frac{d\rho(t)}{dt}  \right|_{t=T_c} 
  &=& \nonumber 
  [\tilde{L}_A+i\tilde{L}_B,\tilde{\rho}(t) (\lambda_A-i\lambda_B)]+h.c. \\
  &=& \nonumber
  -i[(-2\lambda_A \tilde{L}_B),\tilde{\rho}(t)]
  -i[(+2\lambda_B \tilde{L}_A),\tilde{\rho}(t)] \\
  &=&  -i [\tilde{H}_{eff},\tilde{\rho}(t)],
\end{eqnarray}
which contains no decoherence terms. 
The connection to the standard dynamical decoupling can be seen
more transparently if one assumes that $\alpha(t,s)=\theta(|t-s|/T_c)/T_c$
and $H_0=0$. The decoupling conditions then becomes
\begin{equation}
  \int_0^{T_c} ds U_c^\dagger(s) L_{A,B} U_c(s)=\lambda_{A,B},
\end{equation}
which is exactly the standard bang-bang decoupling condition\cite{chen:022343}.

In the following we will derive the non-Markovian master equations 
when continuous or bang-bang decoupling is applied to the system.
Those master equations will be used
in next two sections to simulate the renormalized dynamics.
First consider the case where $L=\sigma_z$. 
Without the control Hamiltonian the master equation is
\begin{eqnarray}
  \frac{d\tilde{\rho}(t)}{dt} 
  &=&  \nonumber
  \left[\sigma_z,\tilde{\rho}(t) \int_0^t ds \alpha^*(t,s) \sigma_z\right]  +
  h.c. . \\
  &=&
  \int_0^t ds \mu(t,s) \left[ \begin{array}{cc}
    0 & -4 \tilde{\rho}_{01} \\
    -4 \tilde{\rho}_{10} & 0 \\
  \end{array} \right],
\end{eqnarray}
where we have decomposed $\alpha(t,s)$ into its real part $\mu(t,s)$ and imaginary
part $\nu(t,s)$ but with the convention $\alpha=\mu-i\nu$.
Note that $\nu(t,s)$ doesn't contribute the decoherence in this case.
According to Ref.[\onlinecite{chen:022343}], a continuous decoupling can be achieved
by using the control Hamiltonian $H_c(t)=a_x(t)\sigma_x$, where we have assumed
$H_0=0$ since we are interested in the quantum memory.
Under such a decoupling the master equation becomes
\begin{equation}
  \frac{d\tilde{\rho}(t)}{dt} = 
  \left[\sigma_z(t),\tilde{\rho}(t) \int_0^t ds \alpha^*(t,s) \sigma_z(s)\right]  +
  h.c. .
\end{equation}
To proceed, one needs to find the associated time evolution operator
\begin{equation}
  U_c(t)=e^{-i\int_0^t du a_x(u) \sigma_x}=
  \left[ \begin{array}{cc}
  \cos(A(t)) & -i \sin(A(t)) \\
  -i\sin(A(t)) & \cos(A(t)) 
  \end{array}\right],
\end{equation}
where $A(t)=\int_0^t du a_x(u)$, and
\begin{equation}
  \sigma_z(s)
  =U^\dagger_c(s)\sigma_z U_c(s)
  =
  \left[ \begin{array}{cc}
  \cos(2A(s)) & -i \sin(2A(s)) \\
  i\sin(2A(s)) & \cos(2A(s)) 
  \end{array}\right].
\end{equation}

It has been shown in Ref.[\onlinecite{chen:022343}] that decoupling condition
can be achieved if $2A(T_c/2)=\pi$ and $2A(T_c/2+s)=\pi+2A(s)$.
Under this condition one has
\begin{widetext}
\begin{eqnarray}
  \int_0^{T_c} dt U_c^\dagger(s)\sigma_z U_c(s) 
  &=& \nonumber
  \int_0^{\frac{T_c}{2}} dt 
  \left[ \begin{array}{cc}
    \cos(2A(s))+\cos(2A(\frac{T_c}{2}+s)) & -i \sin(2A(s))-i \sin(2A(\frac{T_c}{2}+s)) \\
    i\sin(2A(s))+i \sin(2A(\frac{T_c}{2}+s)) & \cos(2A(s)+\cos(2A(\frac{T_c}{2}+s)) 
  \end{array}\right] \\
  &=& 
  \int_0^{\frac{T_c}{2}} dt 
  \left[ \begin{array}{cc}
    \cos(2A(s))-\cos(2A(s)) & -i \sin(2A(s))+i \sin(2A(s)) \\
    i\sin(2A(s))-i \sin(2A(s)) & \cos(2A(s)-\cos(2A(s)) 
  \end{array}\right]=0.
  \end{eqnarray}
\end{widetext}
Hence at the ideal limit where $T_c \rightarrow 0$ the decoherence is totally suppressed, i.e.,
\begin{equation}
 \lim_{T_c \rightarrow 0, N\rightarrow \infty, t=NT_c} \left. 
 \frac{d\tilde{\rho}(t)}{dt} \right|_{t=NT_c}=0.
\end{equation}
However when $T_c$ is finite, the decoupling cannot be perfectly achieved.
The error associated with finite $T_c$ will depend on the correlation function
$\alpha(t,s)$ and the envelope function $a_x(t)$. 
The master equation associate with a finite $T_c$ can be easily derived.
After some algebra we find
\begin{eqnarray}
 \frac{d\tilde{\rho}}{dt} 
&=& \nonumber
  2 \sin(2A(t)) \int_0^t ds \mu(t,s) \sin (2A(s)) \\
&\times& \nonumber
  \left[ \begin{array}{cc}
    \tilde{\rho}_{11}(t)-\tilde{\rho}_{00}(t) &  -\tilde{\rho}_{01}(t)-\tilde{\rho}_{10}(t)  \\
    -\tilde{\rho}_{01}(t)-\tilde{\rho}_{10}(t)  & -\tilde{\rho}_{11}(t)+\tilde{\rho}_{00}(t)  
   \end{array} \right] \\
&+& \nonumber
  2 \sin(2A(t)) \int_0^t ds \nu(t,s) \cos (2A(s)) \\
&\times& 
  \left[ \begin{array}{cc}
    \tilde{\rho}_{01}(t)+\tilde{\rho}_{10}(t) &  -\tilde{\rho}_{00}(t)+\tilde{\rho}_{11}(t)  \\
    -\tilde{\rho}_{00}(t)+\tilde{\rho}_{11}(t)  & -\tilde{\rho}_{01}(t)-\tilde{\rho}_{10}(t)  
   \end{array} \right] ,
\end{eqnarray}
where we have used $\alpha(t,s)=\mu(t,s)-i\nu(t,s)$. Once the envelop function
$a_x(t)$ is identified, this equation can be used to calculate the renormalized dynamics.
Note that under continuous decoupling both $\mu$ and $\nu$ contribute to
the decoherence.

On the other hand for bang-bang control the function $A(t)$ takes the special form
\begin{equation}
  A(t)=\frac{\pi}{2} \sum_{i=0}^{N-1} \Theta(t-i \frac{T_c}{2}).
\end{equation}
It is then straightforward to show that $\sigma_z(s)=f(s)\sigma_z$ where
$f(s)=+1$ for $s \in (2i \frac{T_c}{2},2i+1 \frac{T_c}{2} )$ and $f(s)=-1$ for $s
\in (2i+1 \frac{T_c}{2},2i+2 \frac{T_c}{2})$, $i=1,2,\dots,[\frac{t}{T_c}]$. 
The resulting master equation is
\begin{equation}
  \frac{d\tilde{\rho}(t)}{dt}   = 
  \int_0^t ds \mu(t,s) f(s) \left[ \begin{array}{cc}
    0 & -4 \tilde{\rho}_{01} \\
    -4 \tilde{\rho}_{10} & 0 \\
  \end{array} \right].
\end{equation}
In other words, applying bang-bang control results in the renormalization
of the function $\mu(t,s)$. 
The renormalized $\mu(t,s)$ can be evaluated numerically as follows
\begin{eqnarray}
  & &  
  \int_0^t ds \mu(t,s) f(s) \\
  &=& \nonumber
  \sum_{k=1}^{2N} \int^{k \frac{T_c}{2}}_{(k-1) \frac{T_c}{2}} ds 
  \mu(t,s) (-1)^{k-1} \sigma_z
  +\int_{NT_c}^{t} ds \mu(t,s) \sigma_z .
\end{eqnarray}
We observe that the master equation under bang-bang control takes
a simpler form compared to the one under continuous decoupling.
This suggests that there might be some dynamical effects which cannot
be captured if one only studies the ideal bang-bang limit.

Consider next the case where $L=\sigma_-$.
Without the control Hamiltonian the master equation is
\begin{eqnarray}
  \frac{d\tilde{\rho}(t)}{dt} 
  &=&  \nonumber
  \left[\sigma_-,\tilde{\rho}(t) \int_0^t ds \alpha^*(t,s) \sigma_+\right]  +
  h.c. . \\
  &=& \nonumber
  \int_0^t ds \mu(t,s) 
  \left[ \begin{array}{cc}
    2\tilde{\rho}_{11}(t) & -\tilde{\rho}_{01} \\
    -\tilde{\rho}_{10} & -2\tilde{\rho}_{11}(t) \\
  \end{array} \right] \\
  &+&
  \int_0^t ds \nu(t,s)
  \left[ \begin{array}{cc}
    0 & -i\tilde{\rho}_{01} \\
    +i \tilde{\rho}_{10} & 0 \\
  \end{array} \right].
\end{eqnarray}
Note that in this case both $\mu$ and $\nu$ contribute to the decoherence.
When continuous decoupling is applied the master equation becomes
\begin{eqnarray}
&& \frac{d\rho}{dt} \\
&=& \nonumber
  \left[ \begin{array}{cc}
  -(\tilde{\mu}(t)+\tilde{\mu}^*(t)) \tilde{\rho}_{00}(t) &  -\tilde{\mu}^*(t) \tilde{\rho}_{01}(t) \\
    -\tilde{\mu}(t)  \tilde{\rho}_{10}(t) & -(\tilde{\mu}(t)+\tilde{\mu}^*(t)) \tilde{\rho}_{00}(t)  
   \end{array} \right] \\
&+& \nonumber
  \left[ \begin{array}{cc}
    (-i\tilde{\nu}(t)+ i\tilde{\nu}^*(t)) \tilde{\rho}_{00}(t) & i\tilde{\nu}^*(t) \tilde{\rho}_{01}(t) \\
    -i \tilde{\nu}(t) \tilde{\rho}_{10}(t) &(+i\tilde{\nu}(t)- i\tilde{\nu}^*(t)) \tilde{\rho}_{00}(t)  
   \end{array} \right],
\end{eqnarray}
where
\begin{equation}
\tilde{\mu}(t)=
4 e^{-i2A(t)} \int_0^t ds \mu(t,s) e^{i2A(s)} ,
\end{equation}
and
\begin{equation}
\tilde{\nu}(t)=
4 e^{-i2A(t)} \int_0^t ds \nu(t,s) e^{i2A(s)} .
\end{equation}
While when bang-bang control is applied, the master equation becomes
\begin{eqnarray}
  \frac{d\tilde{\rho}(t)}{dt} 
  &=& \nonumber
  \int_0^t ds \mu(t,s) f(s)
  \left[ \begin{array}{cc}
    2\tilde{\rho}_{11}(t) & -\tilde{\rho}_{01} \\
    -\tilde{\rho}_{10} & -2\tilde{\rho}_{11}(t) \\
  \end{array} \right] \\
  &+&
  \int_0^t ds \nu(t,s) f(s)
  \left[ \begin{array}{cc}
    0 & -i\tilde{\rho}_{01} \\
    +i \tilde{\rho}_{10} & 0 \\
  \end{array} \right].
\end{eqnarray}
In contrast to the case where $L=\sigma_z$, in this case both $\mu$ and $\nu$
contribute to the decoherence.

\section{Ornstein-Uhlenbeck noise}
\label{sec:ou}
In this section we study how the dynamics of a system driven by 
the Ornstein-Uhlenbeck noise is renormalized by the bang-bang and
continuous decoupling pulses.
The Ornstein-Uhlenbeck noise is characterized by the exponential correlation function
\begin{equation}
  \alpha(t,s)=\mu(t,s)=\frac{1}{2\tau}e^{-\frac{|t-s|}{\tau}}.
\end{equation}
This particular form of the correlation function enables us to carry out
analytically most of the calculation. The main purpose is to illustrate 
how the decoherence is renormalized by the bang-bang or continuous
decoupling pulses. 
First consider the case where $L=\sigma_z$.
Using the result in proceeding section, we find that without the control Hamiltonian
the off-diagonal term of the density matrix decays as follows:
\begin{equation}
  \frac{d\tilde{\rho}_{01}}{dt}= -2 \left(1-e^{-t/\tau} \right) \tilde{\rho}_{01}.
\end{equation}
When a control Hamiltonian of the form $H_c(t)=a_x(t)\sigma_x$
is applied and the resulting $A(t)$ has the form $A(t)=\pi t/T_c$,
the renormalized master equation becomes
\begin{eqnarray}
  & & \frac{d\tilde{\rho}(t)}{dt} =
  \frac{2 M_s(t)  \sin\left(2\pi t/T_c\right) }
         {\left(1+\left(2\pi\tau/T_c\right)^2\right)} \\
  &\times& \nonumber
  \left[ \begin{array}{cc}
    \tilde{\rho}_{11}(t)-\tilde{\rho}_{00}(t) &  -\tilde{\rho}_{01}(t)-\tilde{\rho}_{10}(t)  \\
    -\tilde{\rho}_{01}(t)-\tilde{\rho}_{10}(t)  & -\tilde{\rho}_{11}(t)+\tilde{\rho}_{00}(t)  
   \end{array} \right],
\end{eqnarray}
where
\begin{equation}
  M_s(t)=
  e^{-t/\tau}(\frac{2\pi\tau}{T_c})-\frac{2\pi\tau}{T_c} \cos(\frac{2\pi t}{T_c})
  +\sin(\frac{2\pi t}{T_c}).
\end{equation}    
It is important to observe that
\begin{equation}
  \lim_{T_c \rightarrow 0, N\rightarrow \infty, t=NT_c} 
  \frac{2 M_s(t)  \sin\left(2\pi t/T_c\right) } {\left(1+\left(2\pi\tau/T_c\right)^2\right)}=0.
\end{equation}
If a different envelope function $a_x(t)$ is used, the resulting renormalized master equation will have a similar form, but a different time-dependent pre-factor will appear.

On the other hand, if bang-bang decoupling is applied
the renormalized master equation becomes
\begin{equation}  
  \label{eq:OU}
  \frac{d\tilde{\rho}_{01}}{dt}=
  -2 e^{-\frac{t}{\tau}} \left(
  e^{\frac{t}{\tau}}-1+2e^{\frac{T_c}{2\tau}} 
  \frac{1-e^{\frac{NT_c}{\tau}}}{1+e^{\frac{T_c}{\tau}}}
    \right) \tilde{\rho}_{01}.
\end{equation}
Note again that 
\begin{equation}
  \lim_{T_c \rightarrow 0, N\rightarrow \infty, t=NT_c} 
  \frac{1}{2} e^{-\frac{t}{\tau}} \left(
  e^{\frac{t}{\tau}}-1+2e^{\frac{T_c}{2\tau}} 
  \frac{1-e^{\frac{NT_c}{\tau}}}{1+e^{\frac{T_c}{\tau}}}
    \right)=0.
\end{equation}

\begin{figure}
  \includegraphics[scale=0.35]{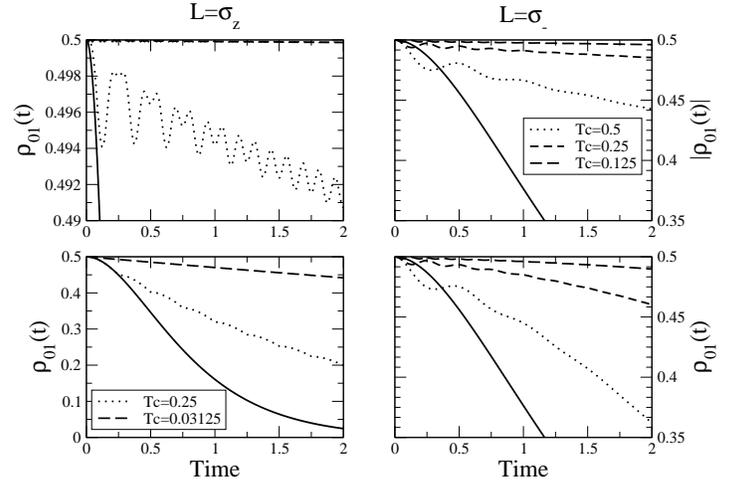}
  \caption{Time evolution of $\rho_{01}(t)$ for a system driven by 
  the Ornstein-Uhlenbeck noise and bang--bang or continuous control. In all figures
  we set $\tau=0.5$. Upper-left: $L=\sigma_z$ and bang-bang control is applied. 
  Bottom-left: $L=\sigma_z$ and continuous control is applied. Right:
  $L=\sigma_z$ and continuous control is applied. In this case $\rho_{01}(t)$ is 
  complex and its absolute value and real part are plotted respectively.}
  \label{fig:OU}
\end{figure}

Consider next the case where $L=\sigma_+$.
In this case the master equation without the control is
\begin{equation}
  \frac{d\tilde{\rho}}{dt}=
  \frac{1}{2}  (1-e^{-t/\tau})
  \left[\begin{array}{cc}
  2\tilde{\rho}_{11}(t) & -2\tilde{\rho}_{01} \\
  -2\tilde{\rho}_{01} &  -2\tilde{\rho}_{11}(t)
  \end{array}\right].
\end{equation}
When the control Hamiltonian $H_c(t)=a_z(t)\sigma_z$ is applied
and $A(t)=\pi t/T_c$, the renormalized master equation is
\begin{eqnarray}
  & & \frac{d \tilde{\rho}(t)}{dt} \\
  &=& \nonumber
    \left[ \begin{array}{cc}
    -(\tilde{\mu}(t)+\tilde{\mu}^*(t)\rho_{00}(t) & -\tilde{\mu}^*(t) \rho_{01}(t) \\
    -\tilde{\mu}(t)\rho_{10}(t) &    (\tilde{\mu}(t)+\tilde{\mu}^*(t)\rho_{00}(t) 
    \end{array}\right],
\end{eqnarray}
where
\begin{equation}
  \tilde{\mu}(t)=2 \frac{1-e^{-t/\tau}e^{-i 2\pi t/T_c}}{1+i 2\pi \tau/T_c}.
\end{equation}
While when bang-bang control is applied, the resulting master equation is
\begin{eqnarray}
  \frac{d\tilde{\rho}(t)}{dt}
  &=& \nonumber
  -\frac{1}{2} e^{-\frac{t}{\tau}} \left(
  e^{\frac{t}{\tau}}-1+2e^{\frac{T_c}{2\tau}} 
  \frac{1-e^{\frac{NT_c}{\tau}}}{1+e^{\frac{T_c}{\tau}}}
    \right) \\
  &\times&   
  \left[ \begin{array}{cc}
    2\tilde{\rho}_{11}(t) & -\tilde{\rho}_{01}(t) \\
    -\tilde{\rho}_{10}(t) & -2\tilde{\rho}_{11}(t) \\
  \end{array} \right].
\end{eqnarray}

It is instructive to simulate numerically the time evolution of $\rho_{01}(t)$ 
under bang-bang and continuous decoupling.
We assume that initially the system is at pure state 
$|\psi\rangle = \frac{1}{\sqrt{2}}(|0\rangle+|1\rangle)$.
In Fig.\ref{fig:OU} (left) we plot $\rho_{01}(t)$ as function of time
for the case where $L=\sigma_z$. The renormalized dynamics
under bang-bang control and continuous decoupling are plotted respectively.
In this case $\rho_{01}(t)$ is always real.
In Fig.\ref{fig:OU} (right) we plot the real part and the absolute value of $\rho_{01}(t)$
for the case where $L=\sigma_-$. We only plot the renormalized dynamics
under continuous decoupling since the dynamics under bang-bang control
is identical to the case where $L=\sigma_z$ with a different overall constant.
 
It is evident that when bang-bang or continuous decoupling is applied
the decoherence is suppressed and better result is achieved for shorter $T_c$.
It is important to note that the efficiency for those two class of decoupling is
different. Continuous decoupling results in a much better suppression.
The continuous decoupling can already suppress the decay of $\tilde{\rho}_{01}$
when $T_c\sim\tau$.
Although we only show one particular realization of the continuous pulse,
numerical evidence shows that continuous pulse always results in 
better suppression for this noise model.
Further optimization on pulse shape is possible and worth more investigation.
It should also be noted that that $\sigma_z$ and $\sigma_-$ coupling to the
environment results in quantitatively different renormalized dynamics.

\section{Spin-boson model}
\label{sec:sb}
In this section we turn our attention to the spin-boson model.
It is known that the zero temperature correlation function of bosonic environment 
can be written as
\begin{equation}
  \alpha(t,s)=\int_0^\infty d\omega J(\omega)
  \left[ \cos(\omega (t-s))-i\sin(\omega (t-s))\right],
\end{equation}
where the spectral density $J(\omega)$ has the form
\begin{equation}
  J(\omega)= \omega^p e^{-\omega/\Lambda_{UV}}.
\end{equation}
When $p=1$ it corresponds to the Ohmic environment, 
when $p>1$ it corresponds to the supra-Ohmic environment, and 
when $p<1$ it corresponds to the sub-Ohmic environment.
For $1/f$ type sub-Ohmic environment, i.e., $p=-1$, 
an infrared cutoff $\Lambda_{IR}$ is also necessary. 


%

\subsection{Ohmic and Supra-Ohmic}


\begin{figure}
  \includegraphics[scale=0.35]{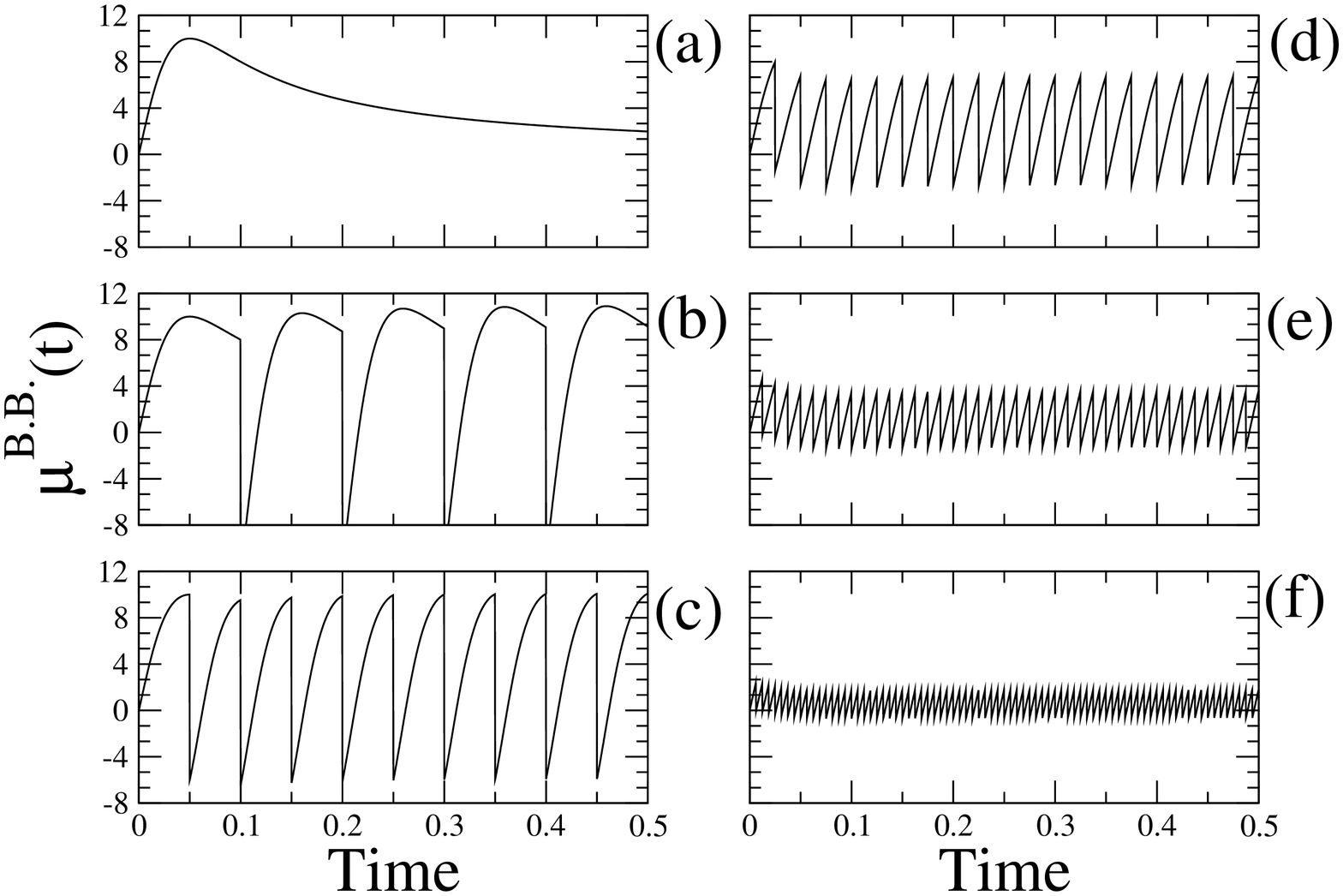}
  \caption{$\mu_1^{B.B}(t)$ as a function of time with an Ohmic environment and
  $\Lambda=20$. (a) Without bang-bang control. (b) $T_c=0.5$.
  (c) $T_c=0.25$. (d) $T_c=0.125$. (e) $T_c=0.0625$.}
  \label{fig:SBmu}
\end{figure}

\begin{figure}
  \includegraphics[scale=0.35]{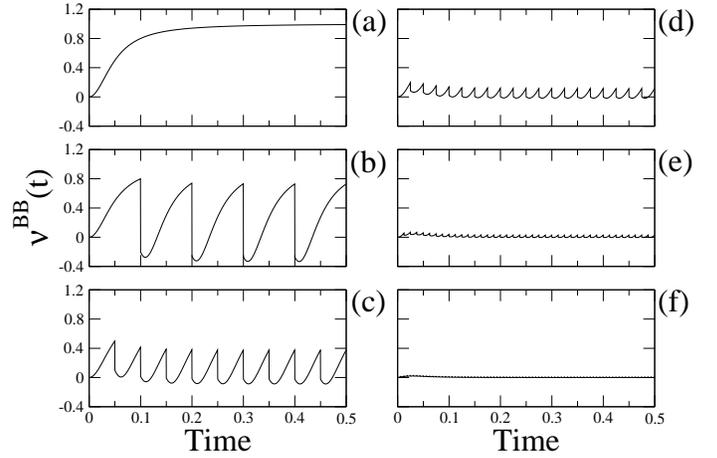}
  \caption{$\nu_1^{B.B}(t)$ as a function of time with an Ohmic environment and
  $\Lambda=20$. (a) Without bang-bang control. (b) $T_c=0.5$
  (c) $T_c=0.25$. (d) $T_c=0.125$. (e) $T_c=0.0625$.}
  \label{fig:SBnu}
\end{figure}

Consider first the Ohmic and supra-Ohmic environment.
It is instructive to exam how $\mu$ and $\nu$ are renormalized by the 
bang-bang control. In Fig.\ref{fig:SBmu} we plot 
$\mu^{B.B.}(t)\equiv \int_0^t ds \mu(t,s) f(s)$ for an Ohmic environment.
It is clear from the figure that when $T_c$ is not small enough,
$\mu^{B.B.}(t)$ might be larger than the original $\mu(t)$ in some temporal regime, which is undesirable. For small enough $T_c$ the $\mu(t)$ is always suppressed.
In Fig.\ref{fig:SBnu} we plot $\nu^{B.B.}\equiv \int_0^t ds \nu(t,s) f(s)$ 
for the same system. We find that $\nu(t)$ is always suppressed by the bang-bang control. Continuous decoupling, on the other hand,  will mix the real and imaginary
part of $\alpha(t,s)$. In some cases this will deteriorate the efficiency of the decoupling.
In the following we present the numerical results for various scenarios.

\subsubsection{$L=\sigma_z$, bang-bang decoupling} 
\begin{figure}
  \includegraphics[scale=0.35]{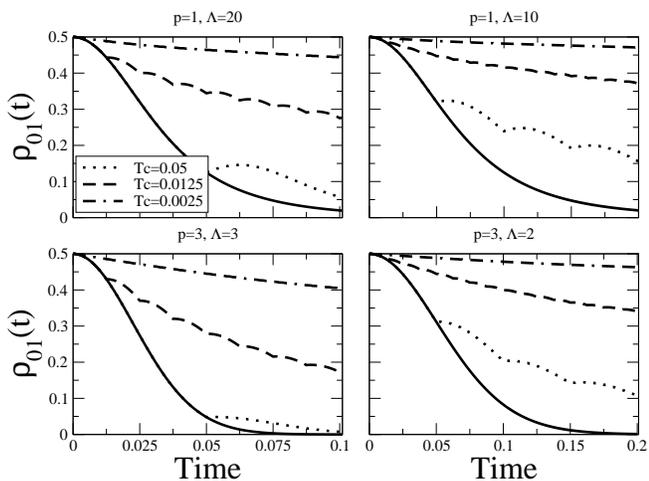}
  \caption{$\tilde{\rho}_{01}(t)$ under bang-bang control for a system coupled to an
  Ohmic ($p=1$) and Supra-Ohmic ($p=3$) environment and $L=\sigma_z$. 
  Solid line represents the free decay without bang-bang control. }
  \label{fig:SBp}
\end{figure}

When $L=\sigma_z$ the master equation at zero temperature without the control
Hamiltonian is reduced to
\begin{equation}
  \frac{d\tilde{\rho}_{01}(t)}{dt}= 
  -4 \int_0^t ds \mu_{p,\Lambda}(t-s) \tilde{\rho}_{01}(t), 
\end{equation}
where
\begin{equation}
  \mu_{p,\Lambda}(t-s)=\int_0^\infty d\omega 
  \omega^p e^{-\omega/\Lambda} \cos(\omega(t-s)).
\end{equation}
When bang-bang control is applied the master equation reads
\begin{equation}
  \frac{d\tilde{\rho}_{01}(t)}{dt}= 
  -4 \int_0^t ds \mu_{p,\Lambda}(t-s) f(s)
  \tilde{\rho}_{01}(t),
\end{equation}
We define $T_2(\Lambda)$ to be the time which satisfies the condition
$\tilde{\rho}_{01}(T_2)=e^{-1}\tilde{\rho}_{01}(0)$. In the Markovian limit
this definition will coincide with the typical definition of $T_2$.
In Fig.\ref{fig:SBp} we plot the time evolution of $\rho_{01}(t)$ under 
bang-bang control for a system coupled to an Ohmic ($p=1$) and 
a supra-Ohmic ($p=3$) environment. 
We have picked two different cut-offs $\Lambda$ for each environment.
It is evident that the free decay is not a simple exponential and
the detail functional form of the decay depends on $p$ and $\Lambda$.
Since the decay is non-exponential, there is no single time scale to which
$T_c$ can be compared. Roughly speaking, however, one can still say that
the decoupling is efficient when $T_c$ is smaller than $T_2(p,\Lambda)$.

\subsubsection{$L=\sigma_z$, continuous decoupling}
\begin{figure}
  \includegraphics[scale=0.35]{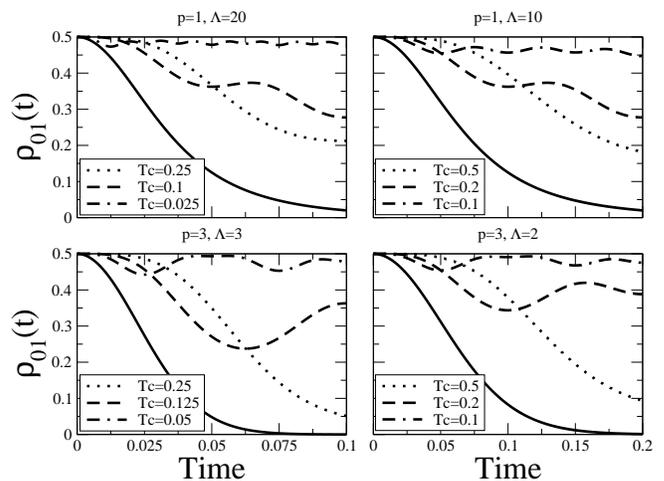}
  \caption{$\tilde{\rho}_{01}(t)$ under continuous decoupling for a system coupled to an
  Ohmic ($p=1$) and Supar-Ohmic ($p=3$) environment and $L=\sigma_z$. 
  Solid line represents  the free decay without decoupling control.}
  \label{fig:SBSzCDD}
\end{figure}

When continuous decoupling is applied the differential equation for
$\tilde{\rho}_{01}(t)$ reads, 
\begin{equation}
  \frac{d\tilde{\rho}_{01}(t)}{dt}=
  -2\sin(\frac{2\pi t}{T_c})\int_0^t ds \mu_{p,\Lambda}(t-s)\sin(\frac{2\pi s}{T_c})
  \tilde{\rho}_{01},
\end{equation}
where we have assumed the same initial condition and $A(t)=\pi t/T_c$.
Note that in this case $\tilde{\rho}_{01}(t)$ remains real during the evolution.
In Fig.\ref{fig:SBSzCDD} we plot the time evolution of $\tilde{\rho}_{01}(t)$ under 
continuous decoupling control for a system coupled to an Ohmic ($p=1$) and 
a supra-Ohmic ($p=3$) environment. Similar to the case where the
system is driven by the Ornstein-Uhlenbeck noise, continuous decoupling is
more efficient for the same $T_c$.

\subsubsection{$L=\sigma_-$, bang-bang decoupling}
\begin{figure}
  \includegraphics[scale=0.35]{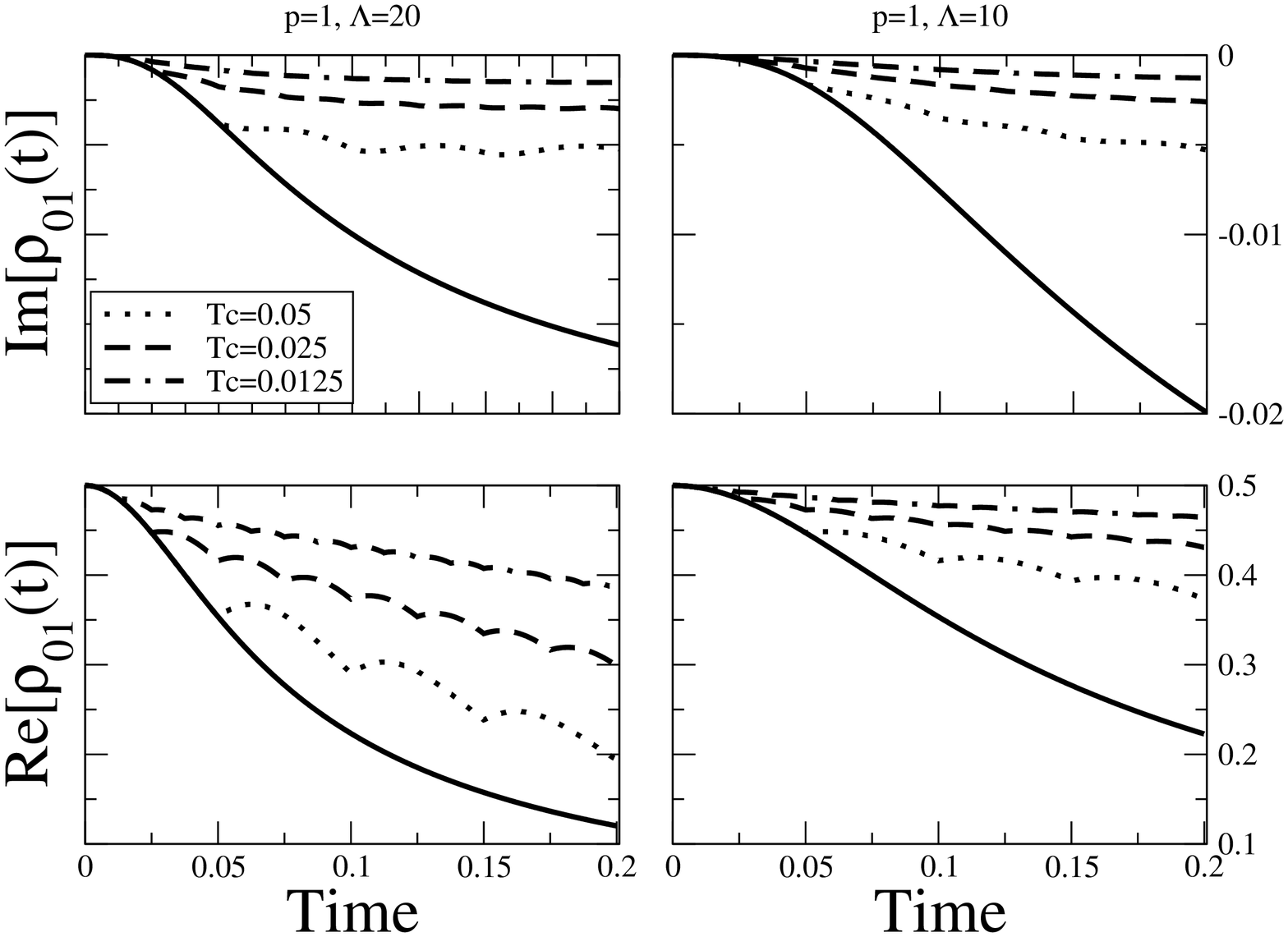}
  \caption{$\tilde{\rho}_{01}(t)$ under bang-bang control for a system coupled to an
  Ohmic ($p=1$) environment and $L=\sigma_-$. 
  Solid line represents the free decay without bang-bang control.}
  \label{fig:SBSM}
\end{figure}

\begin{figure}
  \includegraphics[scale=0.35]{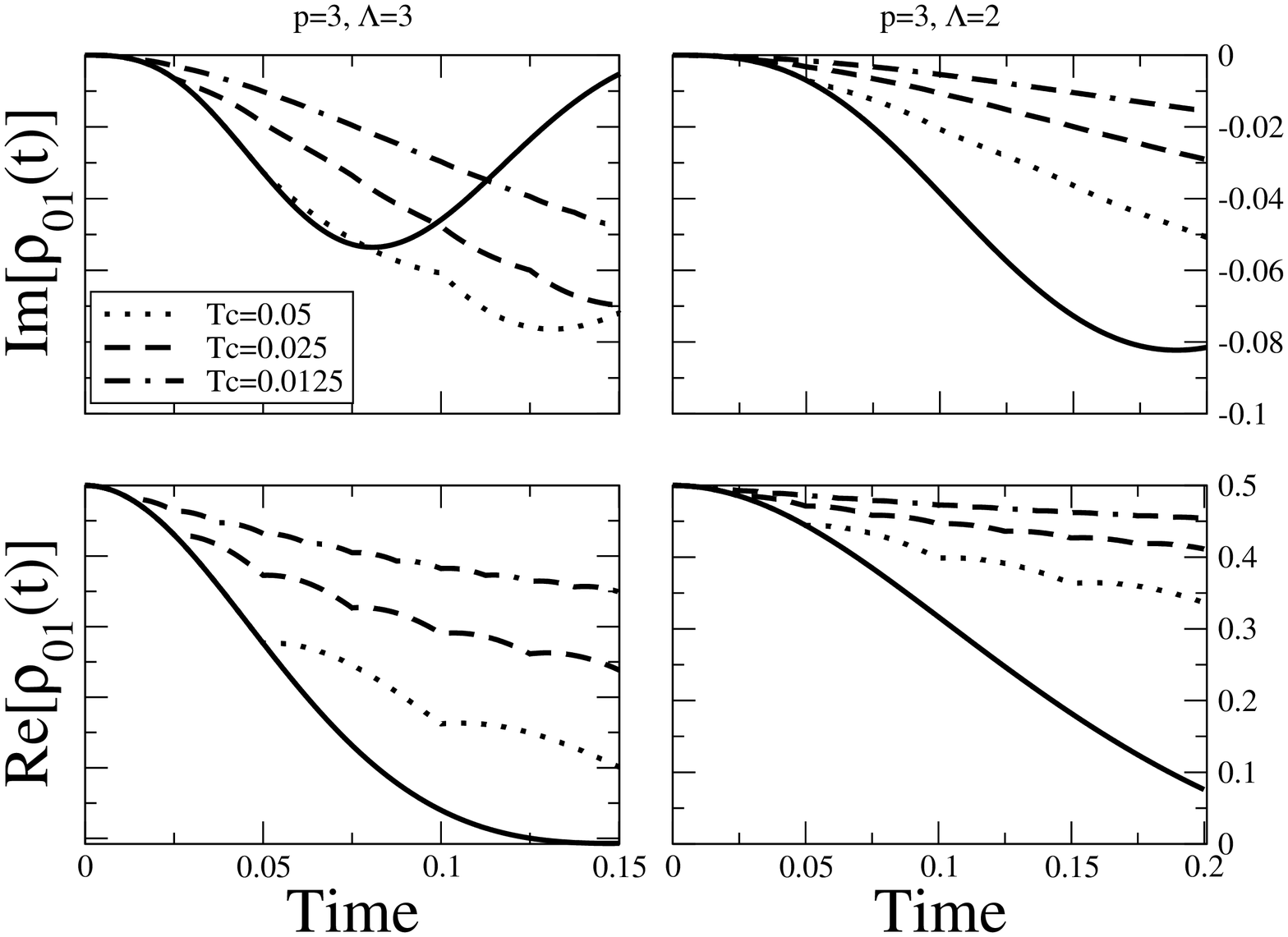}
  \caption{$\tilde{\rho}_{01}(t)$ under bang-bang control for a system coupled to a
  Supar-Ohmic ($p=3$) environment and $L=\sigma_-$. 
  Solid line represents the free decay without bang-bang control.}
  \label{fig:SBSMP3}
\end{figure}

Consider next the case where $L=\sigma_-$.  Without the decoupling pulse 
the differential equation for $\tilde{\rho}_{01}(t)$ reads
\begin{equation}
  \frac{d\tilde{\rho}_{01}(t)}{dt}=
  -\int_0^t ds \left ( \mu_{p,\Lambda}(t,s)+i\nu_{p,\Lambda}(t,s)  \right)
  \tilde{\rho}_{01}(t),
\end{equation}
while when bang-bang control is applied, it becomes
\begin{equation}
  \frac{d\tilde{\rho}_{01}(t)}{dt}=
  -\int_0^t ds \left ( \mu_{p,\Lambda}(t,s)+i\nu_{p,\Lambda}(t,s)  \right) f(s)
  \tilde{\rho}_{01}(t),
\end{equation}
In contrast to the case where $L=\sigma_z$, it is now necessary to evaluate 
the renormalization of both $\mu$ and $\nu$. 
In Fig.\ref{fig:SBSM} (Fig.\ref{fig:SBSMP3}) we plot the 
renormalized dynamics of $\tilde{\rho}_{01}(t)$ under bang-bang control for a system
coupled to an Ohmic (supra-Ohmic) environment. The behavior of the real part
of $\tilde{\rho}_{01}(t)$ is qualitatively similarly to the case where $L=\sigma_z$.
The imaginary part, on the other hand, deviates from zero as time goes.
In most cases it grows monotonically. In some cases, however, 
it shows quasi-oscillation behavior when no decoupling pulse is applied. 
For some range of the $T_c$, the bang-bang control cannot efficiently suppress
the growth of the imaginary part. For small enough $T_c$,  the renormalized dynamics 
of the imaginary part resumes its monotonic increase, with a suppressed increase rate.
It should be noted that the overall efficiency is qualitatively similar to the case
where $L=\sigma_z$  since the decay of the real part of $\tilde{\rho}_{01}$ 
dominates the decoherence.

\subsubsection{$L=\sigma_-$, continuous decoupling}
\begin{figure}
  \includegraphics[scale=0.35]{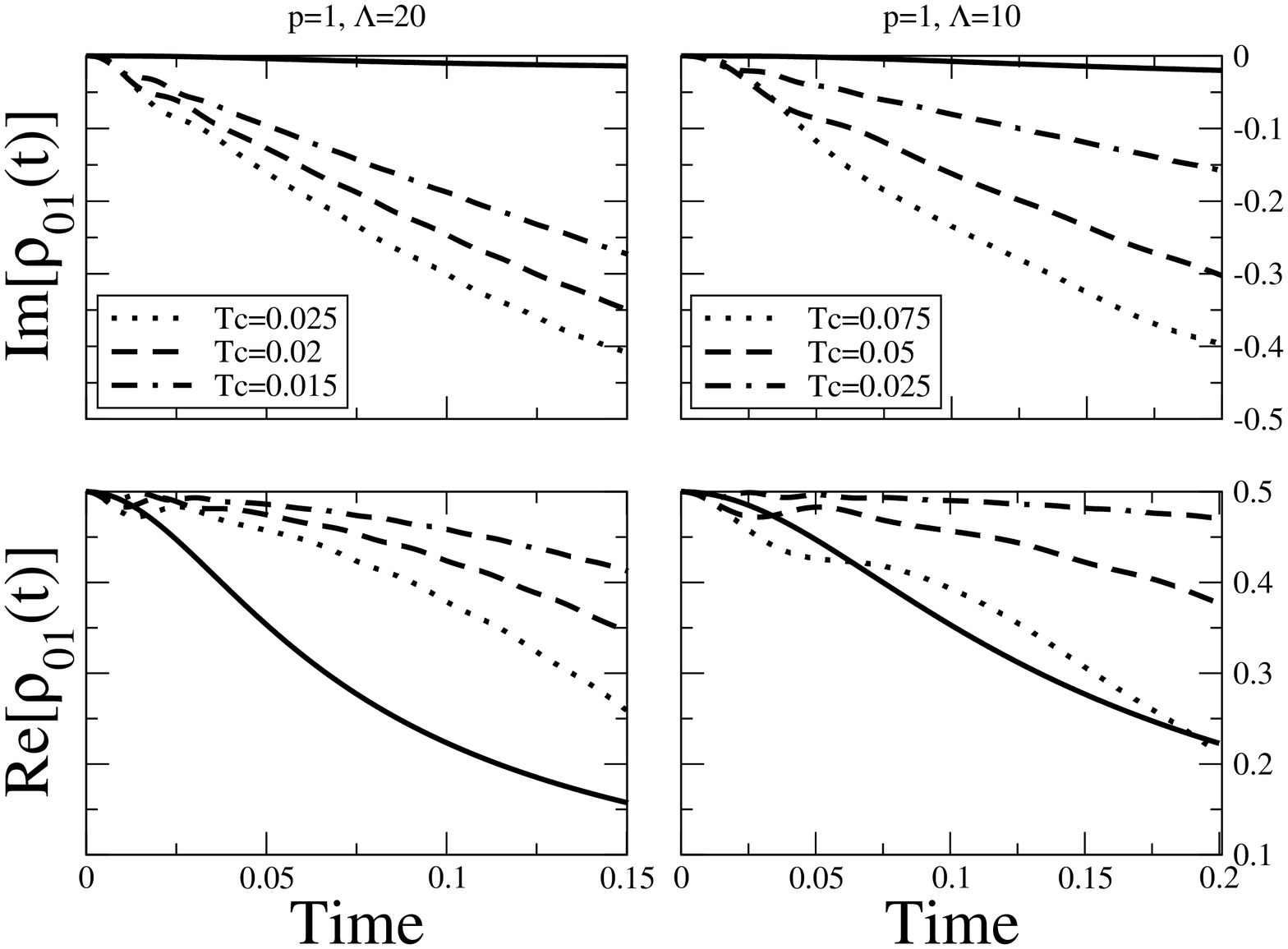}
  \caption{$\tilde{\rho}_{01}(t)$ under continuous control for a system coupled to 
  an Ohmic ($p=1$) environment and $L=\sigma_-$. 
  Solid line represents the free decay without bang-bang control.}
  \label{fig:SBSMCDD}
\end{figure}

\begin{figure}
  \includegraphics[scale=0.35]{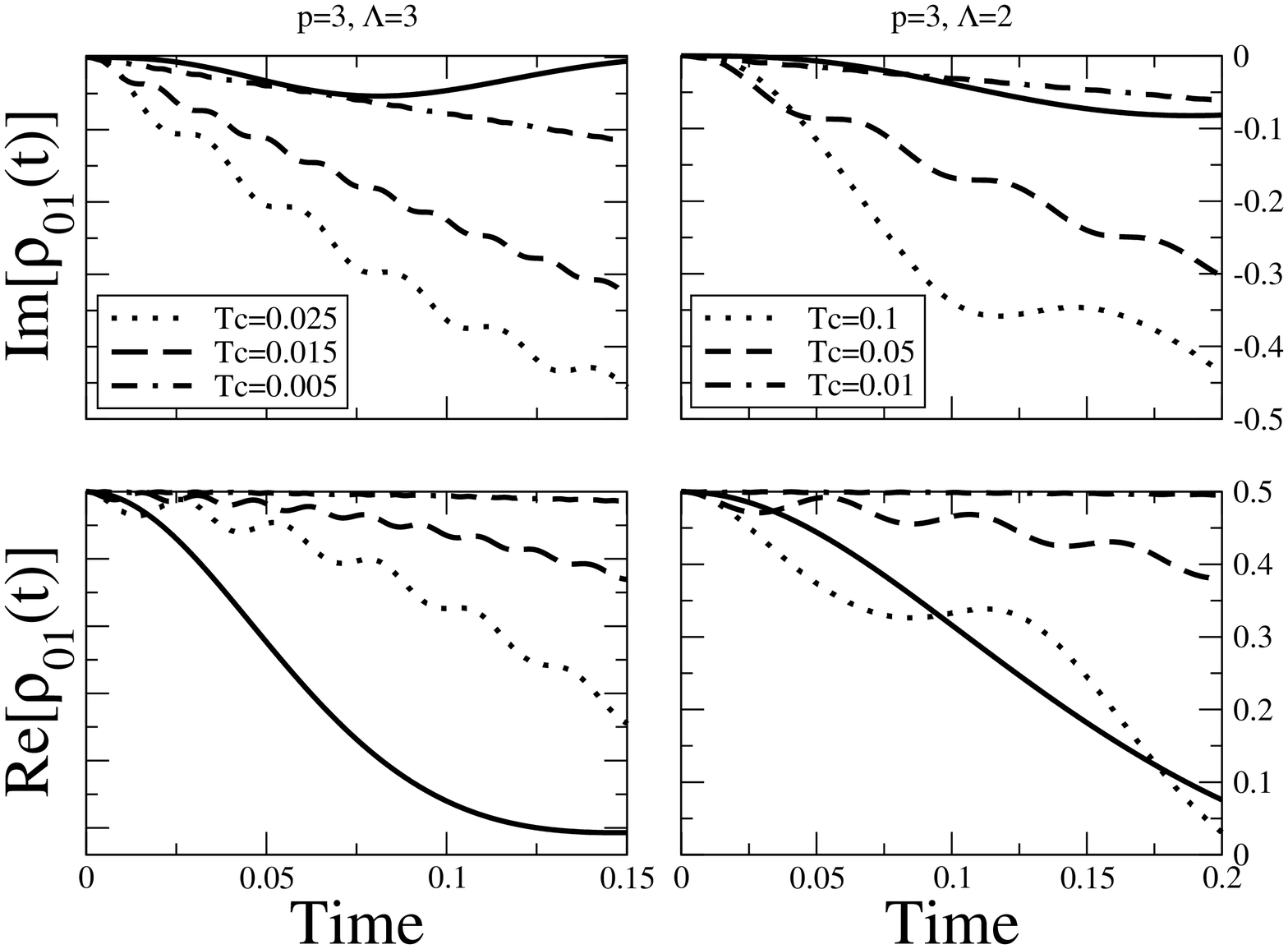}
  \caption{$\tilde{\rho}_{01}(t)$ under continuous control for a system coupled to
  a Supar-Ohmic ($p=3$) environment and $L=\sigma_-$. 
  Solid line represents the free decay without bang-bang control.}
  \label{fig:SBSMCDDP3}
\end{figure}

When continuous decoupling is applied the differential equation for
$\tilde{\rho}_{01}(t)$ reads, 
\begin{equation}
  \frac{d\tilde{\rho}_{01}(t)}{dt}= 
  4 e^{-2iA(t)} \int_0^t ds \left(
  \tilde{\mu}_{p,\Lambda}(t)-i\tilde{\nu}_{p,\Lambda}(t)  \right) e^{2iA(s)}
  \tilde{\rho}_{01}.
\end{equation}
In Fig.\ref{fig:SBSMCDD} (Fig.\ref{fig:SBSMCDDP3}) we plot the 
renormalized dynamics of $\tilde{\rho}_{01}(t)$ under continuous control for a system
coupled to an Ohmic (supra-Ohmic) environment. 
The salient feature to be observed is the dynamics of the imaginary part of
$\tilde{\rho}_{01}(t)$. It only increases slowly for the case of free decay.
When the continuous decoupling is turned on, however, the imaginary part
grows much more rapidly. This behavior results in the deterioration of the efficiency.
The growth of the imaginary part can be suppressed by decreasing the $T_c$.
But a very small $T_c$ is needed to suppress the imaginary part to the same 
level as the one in free decay.
This feature is due to the fact that when continuous decoupling is turned on,
$\mu$ and $\nu$ are not separately renormalized. The decoupling pulse
will mix $\mu$ and $\nu$, resulting a more rapid growth of the imaginary part.

\subsection{Sub-Ohmic}
\begin{figure}
  \includegraphics[scale=0.35]{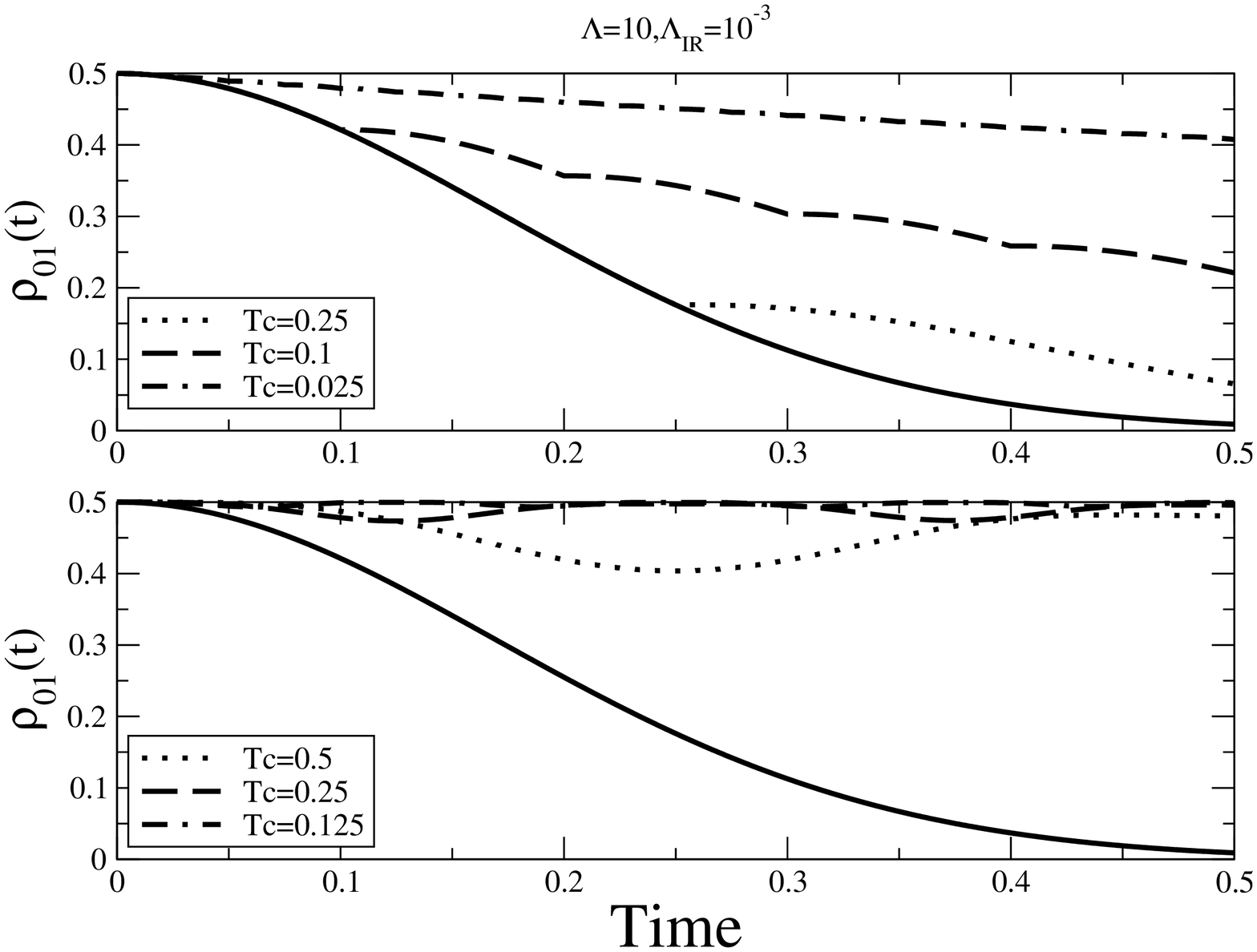}
  \caption{$\tilde{\rho}_{01}(t)$ under bang-bang (upper half) or 
  continuous (bottom half) control for a system coupled to an
  $1/f$-noise environment and $L=\sigma_z$. 
  Solid line represents the free decay without decoupling control.}
  \label{fig:Sz1f}
\end{figure}

\begin{figure}
  \includegraphics[scale=0.35]{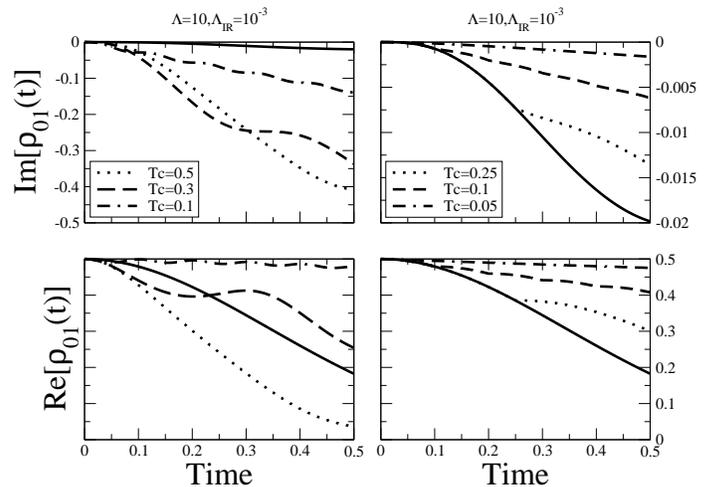}
  \caption{$\tilde{\rho}_{01}(t)$ under bang-bang (right two) or 
  continuous (left two) control for a system coupled to an
  $1/f$-noise environment and $L=\sigma_-$. 
  Solid line represents the free decay without decoupling control.}  \label{fig:Sm1f}
\end{figure}

Now turn our attention to the sub-Ohmic environment. We will focus on the $1/f$ type
environment. For $1/f$ type sub-Ohmic environment the correlation function 
takes the form

\begin{equation}
  \alpha(t,s)=\int_{\Lambda_{IR}}^\infty d\omega 
  \frac{e^{-\frac{\omega}{\Lambda_{UV}}}}{\omega}
  \left[ \cos(\omega (t-s))-i\sin(\omega (t-s))\right],
\end{equation}
where an infrared cutoff is introduced to ensure the convergence.
In Fig.\ref{fig:Sz1f} we plot the renormalized dynamics of $\tilde{\rho}_{01}(t)$ 
for a system with $L=\sigma_z$ under bang-bang or continuous decoupling.
We observe that continuous decoupling is more efficient. 
In Fig.\ref{fig:Sm1f} we plot the renormalized dynamics of $\tilde{\rho}_{01}(t)$ 
for a system with $L=\sigma_-$ under bang-bang or continuous decoupling.
In this case we observe that imaginary begins to grow rapidly once the continuous
decoupling is turned on. This is again due to the mixing of $\mu(t)$ and $\nu(t)$ via
the continuous decoupling. In this scenario the bang-bang control is more efficient.

\section{Summary and discussion}
\label{sec:conclusion}

In summary we have developed a numerical framework to investigate 
the renormalization of the non-Markovian dynamics 
due to the dynamical decoupling pulses. 
The nonconvoluted non-Markovian master equation used in this work
is derived from the non-Markovian quantum trajectories formalism. 
The resulting master equation is written in
a convenient form such that once the decoupling pulses are prescribed, 
the renormalized dynamic can be readily simulated. 
In order to validate
the framework we have performed a comprehensive simulation of
the renormalized dynamics under following scenarios.
We have investigated two different system-environment interactions, 
namely $L=\sigma_z$ and $L=\sigma_-$. 
Three representative environmental spectral densities are used,
including supra-Ohmic, Ohmic, and sub-Ohmic environment. 
Two kinds of decoupling pulses are considered.
One is the typical bang-bang decoupling while 
the other one is the continuous decoupling.
Note also that the period of the decoupling cycle, $T_c$, is finite
in all of the simulations.

The numerical results presented in Section \ref{sec:ou} and  \ref{sec:sb}
clearly demonstrate that the framework is capable of simulate
the renormalization of the dynamics of a system coupled to a general
environment under the action of a general decoupling pulse.
The renormalized dynamics, in turn, determined the true efficiency
of a decoupling pulse.
We observe that different decoupling pulses sometimes give rise to 
qualitatively different renormalized dynamics,
and the difference depends on both the system-environment interaction and 
the environmental correlation function.
Recall that, however, these decoupling pulses are designed 
solely based on the form of the system-environment interaction, 
and they all attain the desired decoupling in the ideal limit.
To better understand the implication of this observation one should notice
that bang-bang decoupling can be viewed as an unphysical limit
of the continuous decoupling\cite{chen:022343}.
Within the framework presented in this work, bang-bang decoupling 
simply corresponds to a very special envelop function. 
On the other hand, only one particular envelop function $A(t)$
is used for continuous decoupling in this work.
In other words, two special envelope function are chosen to carry out
the simulations throughout this work. 
In reality the real envelop function of the pulse, which might be designed
based on other methods, should fall somewhere in between these two limiting cases.
The profound difference in the renormalized dynamics observed 
in this work thus indicates that the real efficiency depends crucially
on the actual shape of the envelop function,
the spectral density of the environment, and the nature of the system-environment
interaction.  For example, our results show that continuous decoupling is
more advantageous in many cases, as can be seem in
Fig.\ref{fig:OU}, Fig.\ref{fig:SBp}, Fig.\ref{fig:SBSzCDD}, and Fig.\ref{fig:Sz1f}.
However when $L=\sigma_-$
the continuous decoupling will induce mixing between the real and imaginary
part of the environment correlation function, making the decoupling less efficient.
Above observation actually leads to the possibility of 
further optimizing the decoupling pulses 
basing on the master equation Eq (\ref{eq:NM_ME_1}) and
using techniques such as pulse shaping\cite{PhysRevB.65.075307}.

The framework developed here can server multiple purposes.
The primary goal is to server as a convenient tool in investigating
the efficiency of a prescribed decoupling pulse. Although we have only
shown the cases where the deterministic, periodic pulses are applied.
The framework can also easily simulate the renormalized dynamics
induced by a random decoupling \cite{viola:037901,chen:022343} or 
a concatenated dynamical decoupling\cite{khodjasteh:180501,khodjasteh:06}.
It can also be used as an alternative starting point to design the decoupling
pulse. One can naturally view the non-Markovian master equation under decoupling
as a dynamical map, whose convergence the ideal fixed point of zero 
system-environment interaction can be studied.
The framework is not limited to a single
two-level system coupled to a general environment. It is also
suitable for studying the renormalized dynamics of a multi-level system.
Such a multi-level system can represent a multi physical or logical qubit system.
When necessary, one can unravel the master equation and use the corresponding
stochastic Schr\"odinger equation, which is numerically advantageous for a large system.
It is also important to note that if eventually a continuous measurement
interpretation of non-Markovian stochastic equations is
developed, this framework can be adapted to study the error correction and
feedback control in the non-Markovian regime.

\begin{acknowledgments}
  We acknowledge the support of the National Science Council in Taiwan through
  Grant No. NSC 94-2112-M-007-018. We thank Dr. K. Shiokawa for
  his critical reading of the manuscript.
\end{acknowledgments}


\begin{thebibliography}{23}
\expandafter\ifx\csname natexlab\endcsname\relax\def\natexlab#1{#1}\fi
\expandafter\ifx\csname bibnamefont\endcsname\relax
  \def\bibnamefont#1{#1}\fi
\expandafter\ifx\csname bibfnamefont\endcsname\relax
  \def\bibfnamefont#1{#1}\fi
\expandafter\ifx\csname citenamefont\endcsname\relax
  \def\citenamefont#1{#1}\fi
\expandafter\ifx\csname url\endcsname\relax
  \def\url#1{\texttt{#1}}\fi
\expandafter\ifx\csname urlprefix\endcsname\relax\def\urlprefix{URL }\fi
\providecommand{\bibinfo}[2]{#2}
\providecommand{\eprint}[2][]{\url{#2}}

\bibitem[{\citenamefont{Viola and Lloyd}(1998)}]{viola:2733}
\bibinfo{author}{\bibfnamefont{L.}~\bibnamefont{Viola}} \bibnamefont{and}
  \bibinfo{author}{\bibfnamefont{S.}~\bibnamefont{Lloyd}},
  \bibinfo{journal}{Phys. Rev. A} \textbf{\bibinfo{volume}{58}},
  \bibinfo{pages}{2733} (\bibinfo{year}{1998}).

\bibitem[{\citenamefont{Zanardi}(1999)}]{PZ99}
\bibinfo{author}{\bibfnamefont{P.}~\bibnamefont{Zanardi}},
  \bibinfo{journal}{Phys. Lett. A} \textbf{\bibinfo{volume}{258}},
  \bibinfo{pages}{77} (\bibinfo{year}{1999}).

\bibitem[{\citenamefont{Viola et~al.}(1999{\natexlab{a}})\citenamefont{Viola,
  Lloyd, and Knill}}]{viola:4888}
\bibinfo{author}{\bibfnamefont{L.}~\bibnamefont{Viola}},
  \bibinfo{author}{\bibfnamefont{S.}~\bibnamefont{Lloyd}}, \bibnamefont{and}
  \bibinfo{author}{\bibfnamefont{E.}~\bibnamefont{Knill}},
  \bibinfo{journal}{Phys. Rev. Lett.} \textbf{\bibinfo{volume}{83}},
  \bibinfo{pages}{4888} (\bibinfo{year}{1999}{\natexlab{a}}).

\bibitem[{\citenamefont{Viola et~al.}(1999{\natexlab{b}})\citenamefont{Viola,
  Knill, and Lloyd}}]{viola:2417}
\bibinfo{author}{\bibfnamefont{L.}~\bibnamefont{Viola}},
  \bibinfo{author}{\bibfnamefont{E.}~\bibnamefont{Knill}}, \bibnamefont{and}
  \bibinfo{author}{\bibfnamefont{S.}~\bibnamefont{Lloyd}},
  \bibinfo{journal}{Phys. Rev. Lett.} \textbf{\bibinfo{volume}{82}},
  \bibinfo{pages}{2417} (\bibinfo{year}{1999}{\natexlab{b}}).

\bibitem[{\citenamefont{Viola and Knill}(2003)}]{viola:037901}
\bibinfo{author}{\bibfnamefont{L.}~\bibnamefont{Viola}} \bibnamefont{and}
  \bibinfo{author}{\bibfnamefont{E.}~\bibnamefont{Knill}},
  \bibinfo{journal}{Phys. Rev. Lett.} \textbf{\bibinfo{volume}{90}},
  \bibinfo{eid}{037901}(\bibinfo{year}{2003}).

\bibitem[{\citenamefont{Chen}(2006)}]{chen:022343}
\bibinfo{author}{\bibfnamefont{P.}~\bibnamefont{Chen}}, \bibinfo{journal}{Phys.
  Rev. A} \textbf{\bibinfo{volume}{73}}, \bibinfo{eid}{022343}
  (\bibinfo{year}{2006}).

\bibitem[{\citenamefont{Byrd and Lidar}(2002)}]{byrd:2002}
\bibinfo{author}{\bibfnamefont{M.~S.} \bibnamefont{Byrd}} \bibnamefont{and}
  \bibinfo{author}{\bibfnamefont{D.~A.} \bibnamefont{Lidar}},
  \bibinfo{journal}{Quantum Inf. Process.} \textbf{\bibinfo{volume}{1}},
  \bibinfo{pages}{19} (\bibinfo{year}{2002}).

\bibitem[{\citenamefont{Viola and Knill}(2005)}]{viola:060502}
\bibinfo{author}{\bibfnamefont{L.}~\bibnamefont{Viola}} \bibnamefont{and}
  \bibinfo{author}{\bibfnamefont{E.}~\bibnamefont{Knill}},
  \bibinfo{journal}{Phys. Rev. Lett.} \textbf{\bibinfo{volume}{94}},
  \bibinfo{eid}{060502} (\bibinfo{year}{2005}).

\bibitem[{\citenamefont{Sengupta and Pryadko}(2005)}]{sengupta:037202}
\bibinfo{author}{\bibfnamefont{P.}~\bibnamefont{Sengupta}} \bibnamefont{and}
  \bibinfo{author}{\bibfnamefont{L.~P.} \bibnamefont{Pryadko}},
  \bibinfo{journal}{Physical Review Letters} \textbf{\bibinfo{volume}{95}},
  \bibinfo{eid}{037202}(\bibinfo{year}{2005}).

\bibitem[{\citenamefont{Khodjasteh and Lidar}(2005)}]{khodjasteh:180501}
\bibinfo{author}{\bibfnamefont{K.}~\bibnamefont{Khodjasteh}} \bibnamefont{and}
  \bibinfo{author}{\bibfnamefont{D.~A.} \bibnamefont{Lidar}},
  \bibinfo{journal}{Phys Rev Lett} \textbf{\bibinfo{volume}{95}},
  \bibinfo{eid}{180501} (\bibinfo{year}{2005}).

\bibitem[{\citenamefont{Facchi et~al.}(2004)\citenamefont{Facchi, Lidar, and
  Pascazio}}]{facchi:032314}
\bibinfo{author}{\bibfnamefont{P.}~\bibnamefont{Facchi}},
  \bibinfo{author}{\bibfnamefont{D.~A.} \bibnamefont{Lidar}}, \bibnamefont{and}
  \bibinfo{author}{\bibfnamefont{S.}~\bibnamefont{Pascazio}},
  \bibinfo{journal}{Phys. Rev. A} \textbf{\bibinfo{volume}{69}},
  \bibinfo{eid}{032314}(\bibinfo{year}{2004}).

\bibitem[{\citenamefont{Shiokawa and Lidar}(2004)}]{shiokawa:030302}
\bibinfo{author}{\bibfnamefont{K.}~\bibnamefont{Shiokawa}} \bibnamefont{and}
  \bibinfo{author}{\bibfnamefont{D.~A.} \bibnamefont{Lidar}},
  \bibinfo{journal}{Phys. Rev. A} \textbf{\bibinfo{volume}{69}},
  \bibinfo{eid}{030302} (\bibinfo{year}{2004}).

\bibitem[{\citenamefont{Faoro and Viola}(2004)}]{faoro:117905}
\bibinfo{author}{\bibfnamefont{L.}~\bibnamefont{Faoro}} \bibnamefont{and}
  \bibinfo{author}{\bibfnamefont{L.}~\bibnamefont{Viola}},
  \bibinfo{journal}{Phys. Rev. Lett.} \textbf{\bibinfo{volume}{92}},
  \bibinfo{eid}{117905}  (\bibinfo{year}{2004}).

\bibitem[{\citenamefont{Shiokawa and Hu}(2005)}]{shiokawa_2005}
\bibinfo{author}{\bibfnamefont{K.}~\bibnamefont{Shiokawa}} \bibnamefont{and}
  \bibinfo{author}{\bibfnamefont{B.~L.} \bibnamefont{Hu}},
  \bibinfo{journal}{e-print quant-ph/0507177}  (\bibinfo{year}{2005}).

\bibitem[{\citenamefont{Caldeira and Leggett}(1983)}]{caldeira:83}
\bibinfo{author}{\bibfnamefont{A.~O.} \bibnamefont{Caldeira}} \bibnamefont{and}
  \bibinfo{author}{\bibfnamefont{A.~J.} \bibnamefont{Leggett}},
  \bibinfo{journal}{Physica A} \textbf{\bibinfo{volume}{121}},
  \bibinfo{pages}{581} (\bibinfo{year}{1983}).

\bibitem[{\citenamefont{Yu et~al.}(1999)\citenamefont{Yu, Diosi, Gisin, and
  Strunz}}]{yu:91}
\bibinfo{author}{\bibfnamefont{T.}~\bibnamefont{Yu}},
  \bibinfo{author}{\bibfnamefont{L.}~\bibnamefont{Diosi}},
  \bibinfo{author}{\bibfnamefont{N.}~\bibnamefont{Gisin}}, \bibnamefont{and}
  \bibinfo{author}{\bibfnamefont{W.~T.} \bibnamefont{Strunz}},
  \bibinfo{journal}{Phys. Rev. A} \textbf{\bibinfo{volume}{60}},
  \bibinfo{pages}{91} (\bibinfo{year}{1999}).

\bibitem[{\citenamefont{de~Vega et~al.}(2005)\citenamefont{de~Vega, Alonso, and
  Gaspard}}]{vega:023812}
\bibinfo{author}{\bibfnamefont{I.}~\bibnamefont{de~Vega}},
  \bibinfo{author}{\bibfnamefont{D.}~\bibnamefont{Alonso}}, \bibnamefont{and}
  \bibinfo{author}{\bibfnamefont{P.}~\bibnamefont{Gaspard}},
  \bibinfo{journal}{Phys. Rev. A} \textbf{\bibinfo{volume}{71}},
  \bibinfo{eid}{023812} (\bibinfo{year}{2005}).

\bibitem[{\citenamefont{Diosi et~al.}(1998)\citenamefont{Diosi, Gisin, and
  Strunz}}]{diosi:1699}
\bibinfo{author}{\bibfnamefont{L.}~\bibnamefont{Diosi}},
  \bibinfo{author}{\bibfnamefont{N.}~\bibnamefont{Gisin}}, \bibnamefont{and}
  \bibinfo{author}{\bibfnamefont{W.~T.} \bibnamefont{Strunz}},
  \bibinfo{journal}{Phys. Rev. A} \textbf{\bibinfo{volume}{58}},
  \bibinfo{pages}{1699} (\bibinfo{year}{1998}).

\bibitem[{\citenamefont{Yu and Eberly}(2002)}]{PhysRevB.66.193306}
\bibinfo{author}{\bibfnamefont{T.}~\bibnamefont{Yu}} \bibnamefont{and}
  \bibinfo{author}{\bibfnamefont{J.~H.} \bibnamefont{Eberly}},
  \bibinfo{journal}{Phys. Rev. B} \textbf{\bibinfo{volume}{66}},
  \bibinfo{pages}{193306} (\bibinfo{year}{2002}).

\bibitem[{\citenamefont{Yu and Eberly}(2003)}]{yu:165322}
\bibinfo{author}{\bibfnamefont{T.}~\bibnamefont{Yu}} \bibnamefont{and}
  \bibinfo{author}{\bibfnamefont{J.~H.} \bibnamefont{Eberly}},
  \bibinfo{journal}{Phys. Rev. B} \textbf{\bibinfo{volume}{68}},
  \bibinfo{eid}{165322}  (\bibinfo{year}{2003}).

\bibitem[{\citenamefont{Magnus}(1954)}]{magnus54}
\bibinfo{author}{\bibfnamefont{W.}~\bibnamefont{Magnus}},
  \bibinfo{journal}{Commun. {P}ure {A}ppl. {M}ath.}
  \textbf{\bibinfo{volume}{7}}, \bibinfo{pages}{649} (\bibinfo{year}{1954}).

\bibitem[{\citenamefont{Piermarocchi et~al.}(2002)\citenamefont{Piermarocchi,
  Chen, Dale, and Sham}}]{PhysRevB.65.075307}
\bibinfo{author}{\bibfnamefont{C.}~\bibnamefont{Piermarocchi}},
  \bibinfo{author}{\bibfnamefont{P.}~\bibnamefont{Chen}},
  \bibinfo{author}{\bibfnamefont{Y.~S.} \bibnamefont{Dale}}, \bibnamefont{and}
  \bibinfo{author}{\bibfnamefont{L.~J.} \bibnamefont{Sham}},
  \bibinfo{journal}{Phys. Rev. B} \textbf{\bibinfo{volume}{65}},
  \bibinfo{pages}{075307} (\bibinfo{year}{2002}).

\bibitem[{\citenamefont{Khodjasteh and Lidar}(2006)}]{khodjasteh:06}
\bibinfo{author}{\bibfnamefont{K.}~\bibnamefont{Khodjasteh}} \bibnamefont{and}
  \bibinfo{author}{\bibfnamefont{D.~A.} \bibnamefont{Lidar}},
  \bibinfo{journal}{e-print quant-ph/0607086}  (\bibinfo{year}{2006}).

\end{thebibliography}

\end{document}